# Non-equilibrium electric double layers at the interface between two electrolytes

S. Hardt and T. Baier

*Institut für Nano- und Mikrofluidik, Center of Smart Interfaces, TU Darmstadt, Petersenstrasse 32, D-64287 Darmstadt, Germany, email: hardt@csi.tu-darmstadt.de*

The charge distribution at the interface between two electrolytes is studied for the case of non-vanishing ion fluxes. The analysis is an extension of the established Verwey-Niessen theory to non-equilibrium situations. Applying matched asymptotic expansions for the region of the electric double layer and the bulk electrolytes, analytical expressions for the ion concentrations and the electrostatic potential are derived. It is found that the electric double layer is surprisingly stable. Even in cases where the applied electric field is so strong that it completely cancels the field of the charge clouds at the interface, two opposing space charge regions are formed. This phenomenon is qualitatively explained using a simple model incorporating ion partitioning and diffusive ion transport. Another notable phenomenon is that the interface acquires a net charge in the presence of an applied electric field, being of relevance for the hydrodynamic stability of interfacial flows.

## I. INTRODUCTION

An electrolyte, being a dielectric medium in which ions are dissolved, can form an interface with a second immiscible electrolyte. Examples for immiscible electrolytes are aqueous two-phase systems [1] or certain ionic liquids in water [2]. Born [3] was probably the first one to note that due to differences in electrostatic energy, ions are generally not equally distributed among the two phases. Rather than that, they partition over the phases and form an electric double layer (EDL), with space charge regions of opposite sign at the two sides of the interface. This situation is reminiscent of EDLs forming at the interface between a solid and a liquid owing to dissociation of surface groups or adsorption of ions, for the first time rigorously studied by Gouy and Chapman [4,5]. The scenario of a back-to-back EDL at the interface between two electrolytes was first analyzed by Verwey and Niessen based on Gouy-Chapman theory [6], and was later extended to include ion adsorption at the interface [7,8]. More elaborate descriptions account for the finite width of the interface [9], or ionic/molecular interactions [10].

In the present work, the structure of an interface between two electrolytes is studied in a non-equilibrium situation. It is analyzed how the double-layer structure changes if the internal electric field, originating from the space-charge regions themselves, is superposed by an external field. At first sight this problem seems to be analogous to determining the charge and field distribution at a heterojunction between two semiconductors. At a pn junction two space charge regions of different signs are formed, and analyzing the response of such a system to an external electric field is a natural problem to study in semiconductor physics. For the latter, hydrodynamic models similar to the one in used in the present work have been applied [11], and some semi-analytical solutions have been derived [12,13]. There are at least two major differences between corresponding models for pn junctions and the approach taken here. First of all, when applying an external voltage to a pn junction, electrons being transported to the other side of the interface recombine with holes. By contrast, in electrolytes where ions are the charge carriers, recombination of anions with cations does usually not occur. Secondly, in semiconductors the spatial extension of the space charge region is often not large compared with the mean-free path of electrons or holes which can be of the order of some ten nanometers at room temperature [14]. From the first argument it follows that in semiconductor devices the mechanism of charge transport is different from electrolytes. From the second argument it can be inferred that the applicability of continuum transport equations for charge carriers is often questionable in semiconductors, whereas due to the very small mean-free path of ions in electrolytes, continuum models can usually be applied without concern.

The purpose of the present work is to extend the Verwey-Niessen theory [6] to non-equilibrium situations with electric current transport across the interface between two electrolytes. The article is structured as follows. In Sec. II, the mathematical model based on the Nernst-Planck equation is presented. In Sec. III an approximate analytical solution based on the method of matched asymptotic expansions is derived. Based on that solution, the electrostatic potential and charge distributions are presented in Sec. IV. The article closes with concluding remarks in Sec. V.

## II. MATHEMATICAL MODEL

It is assumed that a flat interface of negligible thickness between two immiscible electrolytes is located at $X = 0$. A translationally invariant system with no variations in the $Y$ and $Z$ directions, i.e. a one-dimensional problem, is considered. To render the model problem as simple as possible, a system consisting of two ion species, a cation with concentration $P(X)$ and charge $q$, and an anion with concentration $N(X)$ and charge $-q$, is studied. The free energy of the ions (either the Gibbs or the Helmholtz free energy, depending on the constraints applied) is assumed to have the following form

$$F_p(X) = q\Psi(X) \qquad (1)$$
$$F_n(X) = -q\Psi(X) + \mu(X), \qquad (2)$$

where $\Psi$ is the electrostatic potential and $\mu$ the free energy of the cation in the absence of electric fields. According to this model, without electric field only the cation experiences a non-constant free energy. Unequal partitioning of the anion between the two phases is modeled via

$$\mu(X) = \mu_+ \theta(X), \qquad (3)$$

where $\theta$ is the Heaviside function. The problem is analyzed in a model domain $X \in [-L, L]$ in terms of non-dimensional variables. A dimensionless coordinate is defined as $x = X/L$. The contributions to the free energy are nondimensionalized as

$$\Phi = \frac{q}{kT}\Psi, \quad \kappa = \frac{\mu_+}{kT}, \qquad (4)$$

where $k$ is the Boltzmann constant and $T$ the absolute temperature. The concentrations are nondimensionalized using a reference concentration $C$:

$$p = \frac{P}{C}, \quad n = \frac{N}{C}. \qquad (5)$$

Furthermore, it is convenient to introduce dimensionless ion flux densities via

$$j_p = \frac{L}{\nu_p CkT} J_p, \quad j_n = \frac{L}{\nu_n CkT} J_n, \qquad (6)$$

where $\nu_p, \nu_n$ are the electrophoretic mobilities of the cations and anions. The ion concentrations and the electrostatic potential are determined from the Poisson and stationary Nernst-Planck equations. In dimensionless variables, these equations read

$$\lambda^2 \Phi'' - n + p = 0 \qquad (7)$$
$$p' + p\Phi' = -j_p \qquad (8)$$
$$n' - n(\Phi' - \kappa') = j_n, \qquad (9)$$

where a dash denotes the derivative with respect to $x$. The dimensionless parameter $\lambda = \sqrt{\varepsilon kT / L^2 q^2 C}$ measures the ratio of the Debye layer thickness and the spatial extension of the domain. It contains the dielectric permittivity $\varepsilon$ of the electrolyte. The ion flux densities $j_p$ and $j_n$ are constants of integration that are obtained by integrating the original Nernst-Planck equations once. Eqs. (7), (8) and (9) constitute a system of nonlinear ordinary differential equations. The equilibrium problem that can be solved following the approach of Verwey and Niessen [6] is recovered when $j_p = j_n = 0$. In the following section the set of partial differential equations will be solved independently in the two sub-domains with $x > 0$ and $x < 0$, whereupon the two solutions will be matched at the interface. While it needs to be assumed that $\nu_p, \nu_n$ and $\varepsilon$ are constant in each sub-domain, these parameters are allowed to take different values in the two electrolytes.

### III. SOLUTION BASED ON MATCHED ASYMPTOTIC EXPANSIONS

In this section it is described how an approximate analytical solution of Eqs. (7), (8) and (9) can be obtained based on the method of matched asymptotic expansions. For that purpose, the problem is first considered in the interval $x \in (0,1]$ instead of the full interval $x \in [-1,1]$. An apparent choice of the expansion parameter for an asymptotic series is $\lambda$, since the ratio of the Debye layer thickness and the extension of the domain considered is usually a small number. However, expanding the dependent variables of Eqs. (7), (8) and (9) as a power series in $\lambda$ and maintaining the leading-order terms only gives a reasonable approximation in the "bulk" of the electrolyte, i.e. sufficiently far away from the point $x = 0$. Close to $x = 0$, the problem is dominated by the Debye layer and of singularly perturbed structure, meaning that the leading-order terms of the standard asymptotic series do no longer represent a meaningful approximation. Therefore, the problem domain is split up into a boundary-layer domain with $x \approx \lambda$ and a bulk domain with $x \gg \lambda$, and the usual techniques of singular perturbation theory and matched asymptotic expansions are applied [15]. Specifically, in the bulk domain the dependent variables of the Poisson and Nernst-Planck equations are expanded as

$$\Theta(x) = \Theta^{(0)}(x) + \lambda \Theta^{(1)}(x) + ..., \qquad (10)$$

where $\Theta$ stands for any of the three quantities $p, n$ and $\Phi$. In the boundary-layer domain, a scaled variable $\xi = x/\lambda$ is introduced, and $p, n$ and $\Phi$ are regarded as functions of $\xi$. The corresponding asymptotic series is

$$\Theta(\xi) = \Theta^{(0)}(\xi) + \lambda \Theta^{(1)}(\xi) + .... \qquad (11)$$

In other words, in Eq. (10), $x$ is held fixed as $\lambda \to 0$, whereas in Eq. (11) $\xi$ is held fixed. The forms of the boundary-layer and the bulk solutions of the Poisson-Nernst-Planck system have been derived in [16]. A uniformly valid solution is obtained by a suitable matching of the leading-order terms of the two asymptotic expansions. The boundary conditions for the domain $x \in (0,1]$ are

$$\text{for } x \to 0: p = p_+, n = n_+, \Phi = V \qquad (12)$$
$$\text{at } x = 1: p = n = 1, \Phi = 0. \qquad (13)$$

The corresponding solution reads (to $O(\lambda)$ in the expansions)

$$\Phi(x) = \Psi_+^{(0)}(x/\lambda_+) + \Phi_+^{(0)}(x) - \frac{1}{2}\ln(p_+/n_+) - V \quad (14)$$

$$p(x) = p_+ \exp(-\Psi_+^{(0)}(x/\lambda_+) + V) + x(1-\sqrt{p_+n_+}) \quad (15)$$

$$n(x) = n_+ \exp(\Psi_+^{(0)}(x/\lambda_+) - V) + x(1-\sqrt{p_+n_+}), \quad (16)$$

with

$$\Psi_+^{(0)}(x) = V + 2\ln\left(\frac{a\left(1+\frac{1-a}{1+a}\exp(-a\sqrt{2n_+}x)\right)}{1-\frac{1-a}{1+a}\exp(-a\sqrt{2n_+}x)}\right) \quad (17)$$

$$\Phi_+^{(0)}(x) = \left(\frac{1}{2}\ln(p_+/n_+) + V\right)\left(1 - \frac{\ln\left(1-x+\frac{x}{\sqrt{p_+n_+}}\right)}{\ln\left(\frac{1}{\sqrt{p_+n_+}}\right)}\right), \quad (18)$$

where $a = (p_+/n_+)^{1/4}$. Since the analysis should allow considering different concentration levels at $x=1$ and $x=-1$, $C$ in Eqs. (5) and (6) was replaced by $C_+$, and $\lambda$ by $\lambda_+ = \sqrt{\varepsilon_+ kT/L^2 q^2 C_+}$, where $\varepsilon_+$ is the (constant) dielectric permittivity for $x>0$. Correspondingly, for $x<0$, the reference concentration is $C_-$ and $\lambda$ is replaced by $\lambda_- = \sqrt{\varepsilon_- kT/L^2 q^2 C_-}$. For that case the boundary conditions are

$$\text{for } x \to 0: p = p_-, n = n_-, \Phi = U \quad (19)$$
$$\text{at } x = -1: p = n = 1, \Phi = 0 \quad (20)$$

The solution in the interval $x \in [-1,0)$ can then be written as

$$\Phi(x) = \Psi_-^{(0)}(-x/\lambda_-) + \Phi_-^{(0)}(-x) - \frac{1}{2}\ln(p_-/n_-) - U \quad (21)$$

$$p(x) = p_- \exp(-\Psi_-^{(0)}(-x/\lambda_-) + U) - x(1-\sqrt{p_-n_-}) \quad (22)$$

$$n(x) = n_- \exp(\Psi_-^{(0)}(-x/\lambda_-) - U) - x(1-\sqrt{p_-n_-}), \quad (23)$$

where $\Psi_-^{(0)}, \Phi_-^{(0)}$ are obtained from $\Psi_+^{(0)}, \Phi_+^{(0)}$ by replacing $p_+$ with $p_-$ and $n_+$ with $n_-$. The two solutions in $x \in [-1,0)$ and $x \in (0,1]$ need to be matched at $x=0$ to form a solution valid in the full interval $[-1,1]$. By requiring that the electrostatic potential is continuous across the interface, the total voltage drop is obtained as $\varphi = U - V$. Besides $C_+$ and $C_-$, the only parameter that can be controlled externally is the total voltage drop. All other parameters, i.e. $U+V, p_+, n_+, p_-$ and $n_-$ need to be determined through the matching conditions at $x=0$. These five parameters are fixed by the following five conditions

$$p|_{x=0^+} = \tau p|_{x=0^-}, \quad n|_{x=0^+} = \exp(-\kappa)\tau n|_{x=0^-} \quad (24)$$

$$j_p|_{x=0^+} = \tau\sigma j_p|_{x=0^-}, \quad j_n|_{x=0^+} = \tau\sigma j_n|_{x=0^-} \quad (25)$$

$$\Phi'|_{x=0^+} = \Phi'|_{x=0^-}, \quad (26)$$

where $\tau = C_-/C_+$ and $\sigma = \nu_{p-}/\nu_{p+} = \nu_{n-}/\nu_{n+}$. The identical ratios of the electrophoretic mobilities for both types of ions follow if

$$\nu_{p\pm} = \frac{q}{6\pi\eta_\pm r_p}, \quad \nu_{n\pm} = \frac{-q}{6\pi\eta_\pm r_n}, \quad (27)$$

where $\eta_\pm$ are the dynamic viscosities of the two electrolytes and $r_p, r_n$ the hydrodynamic radii of the ions. $f|_{x=0^+}$ ($f|_{x=0^-}$) stands for $\lim_{x\to 0, x>0} f$ ($\lim_{x\to 0, x<0} f$). The factors $\tau$ are due to the different reference concentrations used for negative and positive $x$. The equations express that the ion flux densities and the cation concentrations are continuous at $x=0$. The second identity in Eq. (24) expresses the local equilibrium in the presence of a step change in $\mu$ according to Eq. (3). In addition, since it is assumed that no charges are adsorbed at the interface, the electric field is continuous at $x=0$.

Evaluating Eqs. (24) and (25) and keeping terms up to $O(\lambda)$, the ion concentrations at $x=0$ can be expressed via $U$ and $V$:

$$p_- = \exp(-U)\left(\frac{1+\tau\sigma}{\tau\sigma+\sqrt{\tau^2 r}}\right)^{2\frac{V-U+\ln\tau}{\ln(\tau^2 r)}} \quad (28)$$

$$n_- = \exp(U)\left(\frac{1+\tau\sigma}{\tau\sigma+\sqrt{\tau^2 r}}\right)^{2\frac{V-U+\ln\tau+\ln r}{\ln(\tau^2 r)}} \quad (29)$$

$$p_+ = \tau p_-, \quad n_+ = r\tau n_-, \quad (30)$$

with

$$r = \exp(-\kappa). \quad (31)$$

The two sides of Eq. (26) are evaluated as

$$\Phi'|_{x=0^+} = \frac{\lambda_+\left(1-\sqrt{p_+n_+}\right)\left(2V+\ln\left(\frac{p_+}{n_+}\right)\right) + \sqrt{2}\sqrt{p_+n_+}\left(\sqrt{\frac{p_+}{n_+}}-1\right)\ln(p_+n_+)}{\lambda_+\sqrt{p_+n_+}\ln(p_+n_+)} \quad (32)$$

$$\Phi'\Big|_{x=0^-} = -\frac{\lambda_-\left(1-\sqrt{p_- n_-}\right)\left(2U+\ln\left(\frac{p_-}{n_-}\right)\right)+\sqrt{2}\sqrt{p_- n_-}\left(\sqrt{\frac{p_-}{n_-}}-1\right)\ln(p_- n_-)}{\lambda_-\sqrt{p_- n_-}\ln(p_- n_-)}. \quad (33)$$

---

After some rather tedious algebra, by requiring continuity of the electric field at the interface together with Eqs. (28) to (30), an equation can be derived expressing $V$ as a function of the total voltage drop $\varphi$:

$$V = 2\ln\left(\frac{1}{2}\left(-b+\sqrt{b^2-4c}\right)\right) \quad (34)$$

with

$$b = \exp(-\frac{\varphi}{2})\sqrt{\frac{\varepsilon_+}{\varepsilon_-}}\lambda_-\left(\sqrt{r}-\sigma\right)\left(\sqrt{r}\tau-1\right)(2\varphi+\ln r)$$

$$\cdot\left(\frac{\left(\sqrt{r}+\sigma\right)\tau}{1+\sigma\tau}\right)^{\frac{\varphi+\ln r+\ln\tau}{\ln(r\tau^2)}}\left(\sqrt{2}\left(r+\sqrt{r}\sqrt{\frac{\varepsilon_+}{\varepsilon_-}}\right)(1+\sigma\tau)\ln(r\tau^2)\right)^{-1}$$

(35)

$$c = -\frac{\exp(-\varphi)\left(1+\sqrt{\frac{\varepsilon_+}{\varepsilon_-}}\right)\left(\frac{\left(\sqrt{r}+\sigma\right)\tau}{1+\sigma\tau}\right)^{\frac{2\varphi+\ln r}{\ln(r\tau^2)}}}{\sqrt{r}+\sqrt{\frac{\varepsilon_+}{\varepsilon_-}}}. \quad (36)$$

Via Eqs. (28) to (30), the parameters $p_+, n_+, p_-, n_-$ are then fully determined. Using these expressions in Eqs. (14) to (16) and in Eqs. (21) to (23) results in analytical formulas for the ion concentrations and the electrostatic potential in a system of two immiscible electrolytes in the case of non-vanishing ion currents across the liquid/liquid interface. As will be shown in the next section, in the case of an equilibrium double layer it follows that $\tau = 1/\sqrt{r}$. In this limit the exponents in Eqs. (28) and (29) become singular. However, the expression for the ion concentrations at the interface remain finite, with

$$\lim_{\tau^2 r \to 1} p_- = \sqrt{r}\exp\left(-V+\frac{\sigma\left(V-U-\frac{1}{2}\ln r\right)}{\sigma+\sqrt{r}}\right) \quad (37)$$

$$\lim_{\tau^2 r \to 1} n_- = \frac{1}{\sqrt{r}}\exp\left(V+\frac{\sigma\left(U-V+\frac{1}{2}\ln r\right)}{\sigma+\sqrt{r}}\right). \quad (38)$$

## IV. POTENTIAL AND CHARGE DISTRIBUTION

Before analyzing the non-equilibrium case, the solution derived in the previous section will be compared to the Gouy-Chapman (GC) solution. For this and for the following $\varepsilon_+ = \varepsilon_-$ and $\sigma = 1$ will be assumed. The equilibrium scenario is recovered be setting $j_p = 0$ and $j_n = 0$ in Eqs. (8) and (9). Furthermore, an infinite domain with $L \to \infty$ is considered. In that case, an exact solution is possible, without need to employ matched asymptotic expansions. $U$ and $V$ are uniquely determined by the boundary conditions at $x = \pm 1$ and the matching conditions at $x = 0$. The corresponding equations allow an analytical solution, resulting in

$$U = \ln\left(1+\tanh\left(\frac{\kappa}{4}\right)\right), \quad V = U - \frac{\kappa}{2} \quad (39)$$

The electrostatic potential is obtained as

$$\Phi(x) = \begin{cases} 2\ln\left(\dfrac{1+\exp(\sqrt{2}x/\lambda_-)\tanh(U/4)}{1-\exp(\sqrt{2}x/\lambda_-)\tanh(U/4)}\right) & x < 0 \\ 2\ln\left(\dfrac{1+\exp(-\sqrt{2}x/\lambda_+)\tanh(V/4)}{1-\exp(-\sqrt{2}x/\lambda_+)\tanh(V/4)}\right) & x > 0 \end{cases}. \quad (40)$$

The corresponding charge distributions follow from the Poisson equation, Eq. (7). It is worth noting that unlike in the non-equilibrium case, the concentration ratio $\tau$ cannot take arbitrary values, but is fixed by Eqs. (8) and (9). This can be seen by integrating the equations once and evaluating them at the domain boundaries ($x = \pm 1$). It follows that

$$\tau = \exp\frac{\kappa}{2} \quad (41)$$

The comparison between the GC and the general solution and also the visualization of the charge and potential distributions for the non-equilibrium case will be done in a model domain with $\lambda_- = 0.1$. With $\lambda_-$ being fixed, the Debye layer thickness in the positive half space varies according to $\lambda_+ = \sqrt{\tau}\lambda_-$. Fig. 1 shows the corresponding charge distribution at the interface between the electrolytes for both types of models and different values of $\kappa$. To provide a consistent normalization of the charge for both negative and positive $x$, in this and the following $\rho$ is non-dimensionalized using $C_-$ as a concentration scale. Apparently, both data sets agree well, but with increasing values of $\kappa$ some deviations are found. The deviations result from the fact that the solution presented in the previous section is based on the leading terms of a series expansion in $\lambda_+$ and $\lambda_-$.

When $\lambda_-$ is kept fixed, $\lambda_+$ increases with increasing $\kappa$, making it less justified to neglect higher-order terms in the series expansion. The fact that the transition at $x = 0$ does not appear sharp but exhibits a finite slope is solely due to the resolution chosen for plotting.

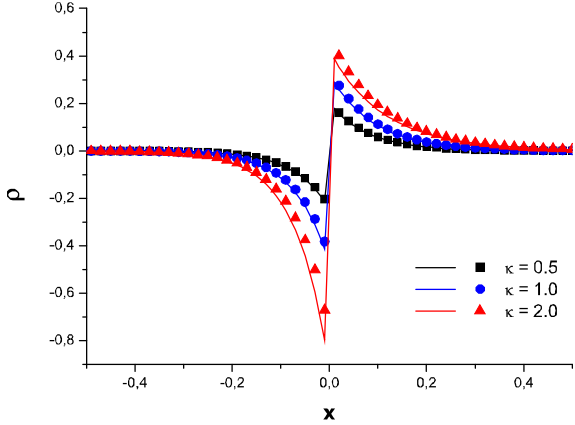

FIG. 1. Comparison of the charge distributions at the interface between two electrolytes obtained with the GC (symbols) and the non-equilibrium model (lines).

In the following, the charge and potential distributions obtained in the non-equilibrium case are analyzed. Fig. 2 shows the dimensionless concentrations (normalized with $C_-$) of positive and negative ions for $V = 5V_{GC}$, $\tau = 1$, and different values of $\kappa$. $V_{GC}$ denotes the value of $V$ as obtained in the Gouy-Chapman model, cf. Eq. (39). The lines represent the results of the non-equilibrium model, the symbols data obtained from a numerical solution of the Nernst-Planck equation. For the latter Eqs. (7) to (9) were solved via the finite element method as implemented in the package COMSOL Multiphysics 3.5a (COMSOL AB, Stockholm, Sweden). The step change of the free energy, Eq. (3), was replaced with a smooth approximation, i.e. the Heaviside function was approximated by the logistic function $\theta_\delta(x) = \frac{1}{2}\left(1 + \tanh(\frac{x}{\delta})\right)$. The parameter $\delta = 10^{-4} \ll \lambda_\pm$ was set small enough such that the results are unaffected by this choice at the length scales of interest. The domain $\{x \mid x \in [-1,1]\}$ was discretized using a non-uniform grid with a spacing varying between $10^{-3}$ at the edges and $10^{-7}$ in the center, using quadratic Lagrange elements for interpolating the trial functions. Dirichlet boundary conditions were applied for the concentrations according to Eqs. (13) and (20), while for the electrostatic potential difference the same value $U - V$ as in the analytical model was chosen.

Apparently, the results of the non-equilibrium model agree very well with the numerical data. Unlike in the equilibrium case, there are now net electromigration and diffusive fluxes passing the interface between the electrolytes. The latter become apparent by the concentration gradients of $p$ and $n$ outside the EDL.

The non-equilibrium model does not rely on any assumptions on the scale of the applied electric field, i.e. it should allow studying fields comparable those created by the EDL itself, of course neglecting additional effects that may come into play such as Joule heating. Fig. 3 shows the dependence of the electrostatic potential and charge distributions on $V$, where the potential curves obtained for positive and negative $x$ have been shifted in such a way that $\Phi(-1) = 0$ and overall continuity is achieved. Variations in $V$ translate to variations of the applied electric field. In Fig. 3 the electrostatic potential is displayed for $\tau = \kappa = 1$ and various values of $V$. The inset of the figure shows the potential in the close vicinity of the interface, for better visualization normalized in such a way that $\Phi(0) = 0$. The corresponding charge distributions are shown in Fig. 4. It becomes apparent that for $V = -10$ and $-5$, the applied electric field is parallel to the field of the EDL, while for $V = 1$ and $5$ both fields are antiparallel.

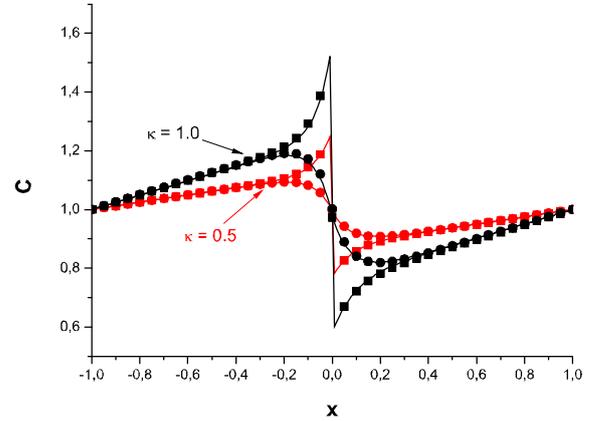

FIG. 2. Concentration profiles of the cations (squares) and anions (circles) for two different values of $\kappa$. The symbols represent numerical results, the corresponding lines the results of the semi-analytical model.

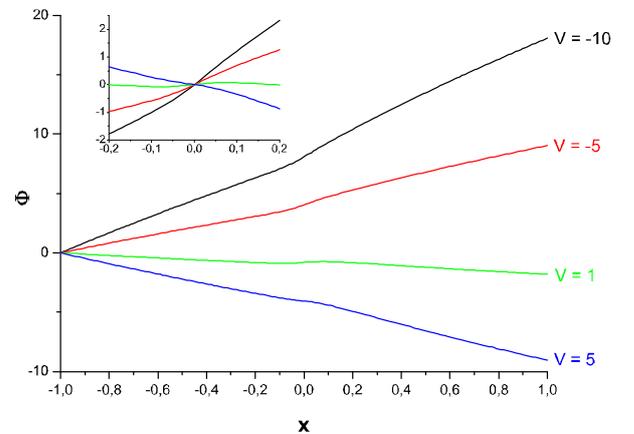

FIG. 3. Electrostatic potential distributions for various values of $V$. The inset shows the potential close to the interface.

Surprisingly, even when the electric field is such that it drags the charge clouds to the right and to the left of the interface apart instead of holding them together (as for $V=5$), the charge distribution across the interface is still qualitatively similar to the Gouy-Chapman case, as apparent from Fig. 4. The structure of the EDL is not fundamentally changed, there is still an excess of positive charges at the right side of the interface and an excess of negative ones at the left.

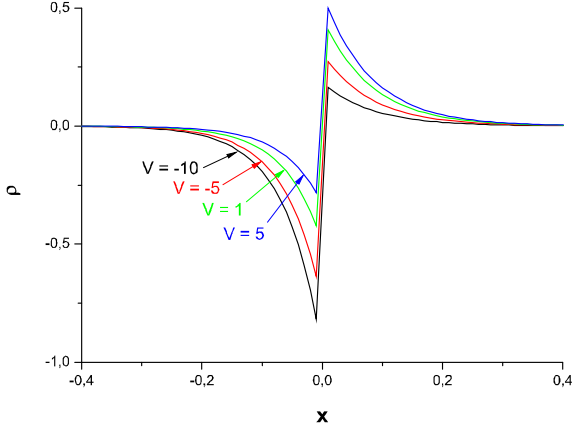

FIG. 4. Charge distribution across the interface for various values of $V$.

To qualitatively explain why the charge clouds stay intact even when the electric field provides no attractive force between the space charge regions, a simplified situation is considered. At a specific value of $V$ a point is reached where the applied field cancels the field due to the charge clouds. Referring to the inset of Fig. 3, this situation will occur at a value lying between 1 and 5. Correspondingly, the case of a vanishing electric field in a finite region $\Omega = \{x \mid x \in [-l, l]\}$ around the interface is studied, where $\Omega$ is large enough to contain the space charge regions and small enough to maintain the assumption of a vanishing electric field. With no electric field, the transport of ions is solely due to diffusion. Since the ion fluxes are independent of $x$ and diffusive fluxes are proportional to the local concentration gradients

$$P'\big|_{x=-l} = P'\big|_{x=l} = -N'\big|_{x=-l} = -N'\big|_{x=l} \equiv K \quad (42)$$

is required. For convenience, here the original and not the nondimensionalized concentration fields of Eq. (5) are used. The fact that the magnitudes of the anion and cation fluxes are identical follows from the assumption of equal transport coefficients for both types of ions. Furthermore, charge neutrality outside of the EDL requires that

$$P(l) = N(l), \; P(-l) = N(-l). \quad (43)$$

The solutions of the one-dimensional diffusion equation are simple linear functions, i.e.

$$P(x) = Kx + P_0, \quad (44)$$

$$N(x) = \begin{cases} -Kx + N_{0-} & x \leq 0 \\ -Kx + N_{0+} & x > 0 \end{cases}. \quad (45)$$

As in Eq. (24), owing to the step change of the free energy of the anions at $x=0$, the discontinuity of the concentration field is

$$\frac{N_{0+}}{N_{0-}} = r. \quad (46)$$

With the help of Eq. (43) it follows that

$$P_0 = 2Kl\frac{r+1}{r-1}, \; N_{0-} = \frac{4Kl}{r-1}. \quad (47)$$

This solution is consistent with overall charge neutrality, i.e. $\int_{-l}^{l}(P-N)dx = 0$. The part of the EDL with positive $x$ carries a charge of

$$Q_+ = q\int_0^l (P-N)dx = -Kl^2 \quad (48)$$

A situation with negligible electric field in the EDL occurs when the field due to the charges in the EDL is compensated by the external field. In the situation considered here, the corresponding external field must be in positive $x$-direction, causing a transport of cations in the same direction. For this reason, $K$ in Eq. (44) has to be negative. Eq. (48) then shows that the charge on the right side of the interface is positive, balanced by a negative charge of equal magnitude on the left side. In other words, the overall distribution of charges remains qualitatively the same as in the equilibrium case without external electric field. The considerations above show that this is possible even without an electric field holding the charge clouds together. Instead, in that case the charge distribution is solely due to ion concentration profiles solving the stationary diffusion equation in combination with a step change of the free energy of the anions at $x=0$.

While in the simple model described above overall charge neutrality of the EDL is maintained, the non-equilibrium model represented by Eqs. (14) to (16) and Eqs. (21) to (23) generally predicts that the interface carries a nonzero charge, i.e. the charges at both sides of the interface do not cancel. Regarding $V$ and $U$ as parameters, an analytical formula for the indefinite integral of $p(x) - n(x)$ can be derived. However, since $V$ and $U$ are connected by a quite complex relationship (Eq. (34)) the analytical expression for $\int_{-1}^{1}(p-n)dx$ is not reproduced here, but the integral is evaluated numerically.

The results for the net charge of the interface are displayed in Fig. 5. The concentration ratio $\tau$ was fixed to the value obtained for the Gouy-Chapman case (Eq. (41)), and $V$ and $\kappa$ were varied. When $V = V_{GC}$, the equilibrium EDL is recovered. For that case the Gouy-Chapman model predicts a vanishing net charge, i.e. a perfect balance of the charges at both sides of the interface. As apparent from Fig. 5, the non-equilibrium model reproduces that result. Apart from that, the net charge displays a linear variation with $V$, indicated by the excellent modeling of the data via linear functions. A similar linear behavior is also found for other values of the concentration ratio $\tau$.

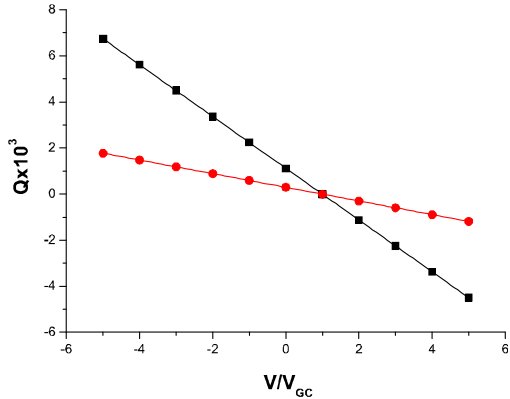

FIG. 5. Net charge created at the interface as a function of $V$ for $\kappa = 1$ (squares) and $\kappa = 0.5$ (circles). The lines represent linear fits to the data.

The results displayed in Fig. 5 show that interfaces between two electrolytes generally acquire a net charge when exposing them to a normal electric field. A comparison with the Gouy-Chapman case shows that this charge is significant: For $\kappa = 1$, at equilibrium each side of the interface carries a charge with a magnitude of $\approx 3.1 \cdot 10^{-2}$, and at $V = 4V_{GC}$ the applied field generates about 10% of that value at the interface. At the interface between two conductive media without a step change of the ionic free energy, a net charge only accumulates if the conductivities of the two media are different. However, when a free-energy jump exists, the point of zero charge is found in a situation (the equilibrium configuration) where the asymptotic ion concentrations in the two media are different. In general this means that the interfacial charge vanishes in a situation where the conductivities of the two electrolytes are different. Vice versa, the interface acquires a net charge in the case of equal conductivities.

The charging is certainly of relevance for the fluid dynamics of liquid/liquid interfaces. The electrostatic force per unit area of the interface is equal to the charge per unit area times the electric field, resulting in significant forces that may trigger hydrodynamic instabilities [17-19]. In addition, if the electric field has components tangential to the interface, an electroosmotic flow may be created.

## V. SUMMARY AND CONCLUSIONS

An analytical model describing the electrostatic potential and ionic charge distribution across an interface between two electrolytes with an uneven partitioning of one of the ion species has been developed. The model addresses non-equilibrium scenarios with an electric field applied normal to the interface, creating ionic currents. It is based on the Nernst-Planck equation, solved by the leading-order terms of matched asymptotic expansions in a small parameter, the ratio between the electric double layer thickness and the extension of the domain of interest. The model shows that even when the applied electric field exceeds the strength of the field produced by the EDL and causes a repulsion between the charge clouds, the qualitative structure of the charge distribution is still maintained. This can be explained in a scenario with a vanishing electric field at the interface, where the diffusive ion fluxes in combination with the boundary condition at the interface result in a charge separation qualitatively similar to the equilibrium case. Furthermore, interfaces between electrolytes can carry a significant charge in the presence of a normal electric field. This may cause hydrodynamic instabilities in liquid/liquid systems, for example being relevant in microfluidic devices.